\documentclass[sigplan,10pt,screen]{acmart}

\setcopyright{none}
\renewcommand\footnotetextcopyrightpermission[1]{}

\acmYear{2026}

\usepackage{amsmath}

\usepackage{amssymb}
\usepackage{booktabs}
\usepackage{subcaption}
\usepackage{algorithm}
\usepackage{algpseudocode}
\usepackage{float} 
\usepackage{array}
\usepackage{tabularx}

\DeclareUnicodeCharacter{00A0}{ }

\title{Asynchronous Verified Semantic Caching for Tiered LLM Architectures}

\author{Asmit Kumar Singh}
\affiliation{
  \institution{Apple}
  \country{USA}
}
\email{asingh77@apple.com}

\author{Haozhe Wang}
\affiliation{
  \institution{Apple}
  \country{USA}
}
\email{haozhe_wang@apple.com}

\author{Santosh Attaluri}
\affiliation{
  \institution{Apple}
  \country{USA}
}
\email{lattaluri@apple.com}

\author{Tak Chiam}
\affiliation{
  \institution{Apple}
  \country{USA}
}
\email{t_chiam@apple.com}

\author{Weihua Zhu}
\affiliation{
  \institution{Apple}
  \country{USA}
}
\email{weihua_zhu@apple.com}

\renewcommand{\shortauthors}{Singh et al.}

\begin{document}

\begin{abstract}
Large language models (LLMs) now sit in the critical path of search, assistance, and agentic workflows, making semantic caching essential for reducing inference cost and latency. Production deployments typically use a tiered static-dynamic design: a static cache of curated, offline vetted responses mined from logs, backed by a dynamic cache populated online. In practice, both tiers are commonly governed by a single embedding similarity threshold, which induces a hard tradeoff: conservative thresholds miss safe reuse opportunities, while aggressive thresholds risk serving semantically incorrect responses. We introduce \textbf{Krites}, an asynchronous, LLM-judged caching policy that expands static coverage without changing serving decisions. On the critical path, Krites behaves exactly like a standard static threshold policy. When the nearest static neighbor of the prompt falls just below the static threshold, Krites asynchronously invokes an LLM judge to verify whether the static response is acceptable for the new prompt. Approved matches are promoted into the dynamic cache, allowing future repeats and paraphrases to reuse curated static answers and expanding static reach over time. In trace-driven simulations on conversational and search workloads, Krites increases the fraction of requests served with curated static answers (direct static hits plus verified promotions) by up to $\textbf{3.9}$ times for conversational traffic and search-style queries relative to tuned baselines, with unchanged critical path latency.
\end{abstract}


\maketitle

\medskip

\section{Introduction}

Large language models (LLMs) have transitioned from research prototypes to core infrastructure for conversational assistants, search engines, productivity tools, and code assistants \citep{zhao2023surveyllm,achiam2023gpt4technical}. Their deployment is constrained by a familiar systems triad of cost, latency, and quality: each prompt triggers multiple expensive forward passes through a large transformer, and inference often dominates the budget \citep{kwon2023pagedattention,xiong2024services}. This pressure is amplified in modern \emph{agentic} stacks that orchestrate workflows with multiple steps(retrieval, planning, tool use, reflective revision) across multiple model calls \citep{wang2023surveyagents}, including emerging LLM-centric search stacks \citep{wang2024largesearch,liu2024searchllm}. For such workloads, even modest reductions in backend invocations can yield significant gains in cost and tail latency.

Caching is the standard systems lever for reuse. Information retrieval (IR) systems have long applied result caching to avoid recomputing expensive ranking pipelines \citep{Fagni2006Caching,BaezaYates2008Design}. At large scale, production systems are often tiered: a \emph{static} cache populated offline from historical logs, backed by a \emph{dynamic} cache populated online to absorb tail traffic and short-term trends \citep{Fagni2006Caching,BaezaYates2008Design,Mele2020Topical}. Static entries are often curated by heavier offline pipelines (larger models, safety review, human evaluation), so each additional \emph{static origin} hit can carry\textit{ outsized value} in stability and reliability relative to serving an online generated answer.

Semantic caching extends caching to LLM workloads by relaxing exact key matching. Instead of requiring identical strings, a semantic cache embeds each prompt into a vector space and retrieves nearest neighbors from a vector database \citep{bang2023gptcache,pan2024vectordb}. If similarity exceeds a threshold, the cache reuses the cached response and avoids invoking the LLM backend. Semantic caching can substantially reduce latency on workloads with recurring intents and paraphrases \citep{bang2023gptcache,zhu2024efficientembeddings}, and is especially effective for short to medium prompts such as search queries and assistant calls \citep{schroeder2025vcache}.

\begin{figure*}[t]
  \centering
  \begin{subfigure}[b]{0.50\textwidth}
    \centering
    \includegraphics[width=\linewidth]{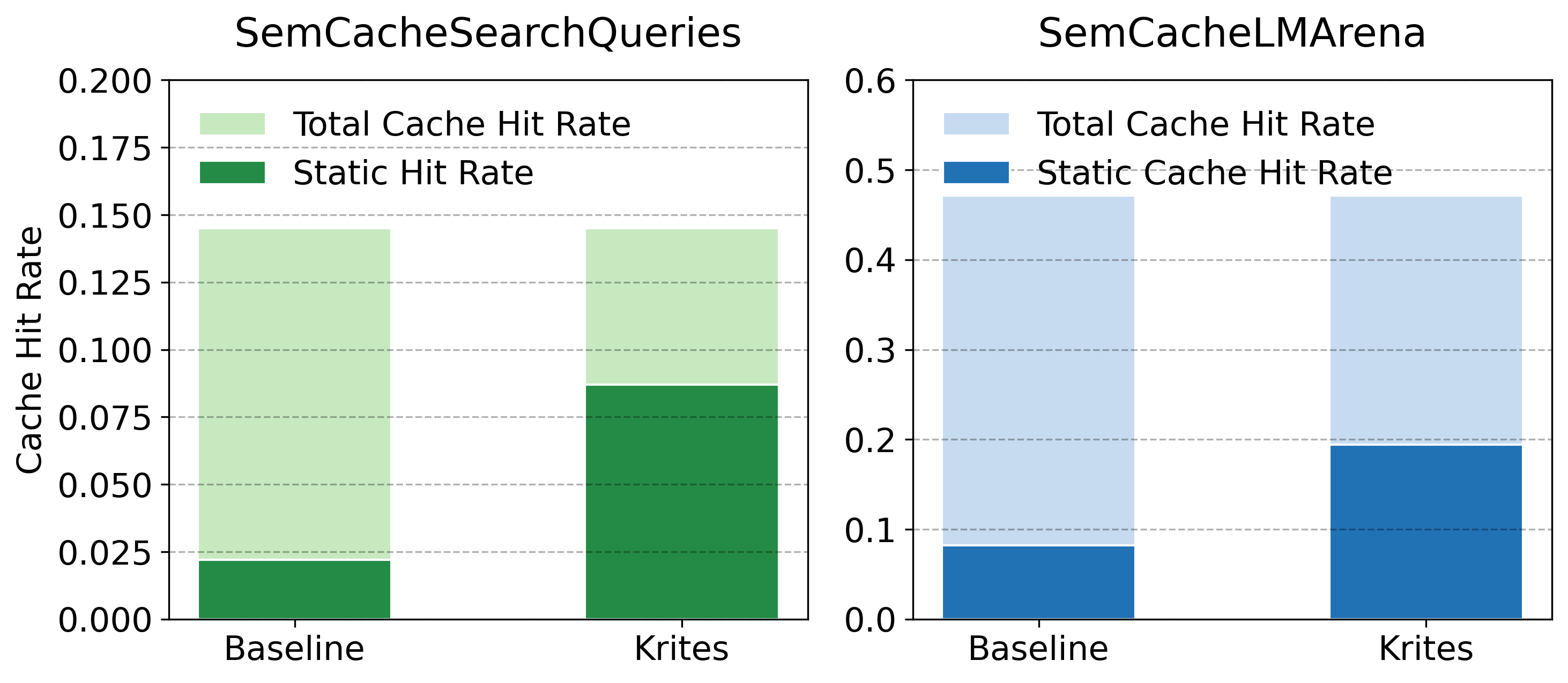}
    \caption{Cache hit rate comparisons.}
    \label{fig:hit-composition}
  \end{subfigure}
  \hfill
  \begin{subfigure}[b]{0.35\textwidth}
    \centering
    \includegraphics[width=\linewidth]{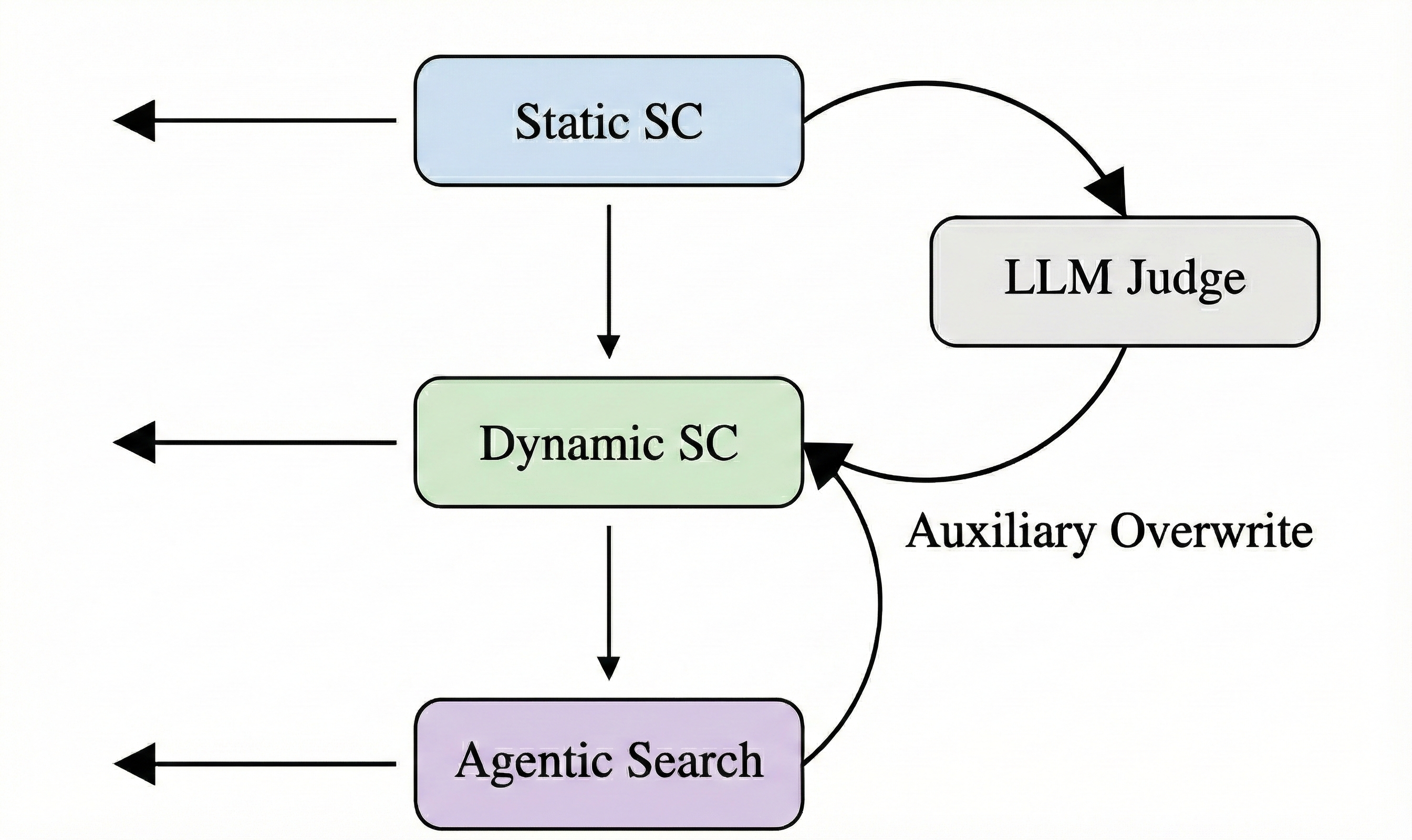}
    \caption{Krites policy.}
    \label{fig:krites-policy}
  \end{subfigure}

  \caption{(a) Hit rate composition for the baseline static threshold policy versus Krites. The overall cache hit rate (total bar height) remains identical. However, Krites increases the proportion of requests served with curated static-origin answers (direct static hits plus verified promotions). (b) Krites couples the static and dynamic tiers via an off-path LLM judge and auxiliary overwrite.}
  \Description{Left: stacked bars comparing baseline vs. Krites hit composition. Right: two-tier cache diagram with off-path LLM judge and auxiliary overwrite into the dynamic cache.}
  \label{fig:banner-figure}
\end{figure*}
Despite the tiered static--dynamic architecture, most semantic caches are still governed by a single global similarity threshold applied uniformly across requests. Let $q$ be an incoming prompt and $h$ its nearest neighbor in a cache under cosine similarity. The standard policy is
\[
\text{hit if } \operatorname{sim}(q, h) \ge \tau, \quad \text{miss otherwise},
\]
with separate constants $\tau_{\text{static}}$ and $\tau_{\text{dynamic}}$ for the two tiers. This design induces a direct tradeoff between hit rate and error rate: lowering $\tau$ recovers more reuse opportunities but increases false hits, while raising $\tau$ reduces errors but yields more backend calls \citep{bang2023gptcache,rekabsaz2017threshold}.

Recent work on verified semantic caching makes this limitation precise. vCache \citep{schroeder2025vcache} introduces benchmarks for semantic caching and shows that similarity distributions for correct and incorrect hits overlap heavily. In practice, this creates a similarity \emph{grey zone} where embedding geometry alone cannot reliably separate paraphrases from distinct intents. For example, the following pairs can fall below conservative thresholds while remaining clearly interchangeable:
\begin{center}
{\small
\setlength{\tabcolsep}{4pt}
\begin{tabularx}{\linewidth}{@{}>{\raggedright\arraybackslash}X >{\raggedright\arraybackslash}X@{}}
\toprule
Incoming query $q$ & Cached neighbor $h$ \\
\midrule
What's the word on my dog having honey? & Can my dog have honey? \\
Was anyone successful in the lottery last night? & Anybody win the lottery last night? \\
\bottomrule
\end{tabularx}
}
\end{center}
This suggests a natural systems question: can we preserve a conservative, latency-friendly threshold policy on the critical path, but use a high precision equivalence check to safely recover static misses from the grey zone?

LLM-as-a-judge provides such an equivalence check. When prompted with explicit rubrics and candidate pairs, LLM judges can achieve high agreement with semantic equivalence labels across a range of tasks \citep{zheng2023judging,Liu2024LLMJudgeSurvey,Tan2024JudgeBench}, and pairwise equivalence questions with explicit criteria are particularly tractable \citep{Zeng2025Equivalence}. However, placing an LLM judge directly on the serving path adds latency and compute to borderline requests, eroding the main benefit of caching for interactive workloads.

This paper proposes \textbf{Krites}, an asynchronous, verified semantic caching policy for tiered static--dynamic architectures. Krites preserves the standard GPTCache-style threshold policy on the serving path, but when a request misses the static tier while its nearest static neighbor falls in a grey zone, Krites schedules an off path LLM judge that decides whether the \emph{static response} is acceptable for the new prompt. Approved pairs trigger an \emph{auxiliary overwrite} that upserts the static response into the dynamic cache under the new key, turning the dynamic cache into a mutable pointer layer over static answers (Figure~\ref{fig:banner-figure}b).

Our contribution is the Krites policy: an asynchronous LLM-judged semantic caching mechanism that decouples serving from verification and uses an auxiliary overwrite to promote validated static responses into the dynamic cache, effectively turning the dynamic cache into a mutable pointer layer over the static cache. In trace driven simulations on SemCacheLMArena and SemCacheSearchQueries, Krites increases the fraction of requests served with static origin (curated) answers by up to 136 percent on conversational workloads and by 290 percent on search style workloads at fixed cache error rate and with no increase in critical-path latency compared to tuned static threshold baselines.

Crucially, in production environments like enterprise search or medical-domain assistants, replacing a dynamically generated response with an offline vetted, high quality ``gold'' static answer strictly elevates safety and reliability. As illustrated in Figure~\ref{fig:banner-figure}a, Krites transforms the composition of cache hits: drastically expanding the fraction of traffic served directly from the curated static tier. By doing so, Krites safely unlocks the outsized value of curated responses that would otherwise remain stranded in the similarity grey zone.

\section{Background and system model}
\label{sec:system-model}

We now formalize the tiered semantic caching setting and the baseline policy that Krites augments.

\subsection{Vector embeddings and semantic search}

Semantic caches rely on an embedding model $\Phi$ that maps a discrete prompt $q$ into a vector in $\mathbb{R}^d$:
\begin{equation}
v_q = \Phi(q).
\end{equation}
Similarity between two prompts $q$ and $x$ is measured in this space, typically with cosine similarity:
\begin{equation}
\operatorname{sim}(q, x) = \frac{v_q \cdot v_x}{\|v_q\| \, \|v_x\|}.
\end{equation}
A semantic cache assumes that when $\operatorname{sim}(q, x)$ is close to $1$, the cached response for $x$ is likely a valid response for $q$ \citep{bang2023gptcache,pan2024vectordb}.

\subsection{Tiered static-dynamic architecture}

In large scale web search and assistant systems, caching is typically tiered to balance quality and freshness \citep{Fagni2006Caching,BaezaYates2008Design,Mele2020Topical}. We adopt a two tier semantic cache in front of an agentic LLM backend.

\subsubsection{Static tier $C_{\text{static}}$}

The static tier is a read only cache populated by offline pipelines.
\begin{itemize}
    \item \textbf{Population.} Historical query logs are mined to select head and torso prompts that occur frequently.
    \item \textbf{Quality.} For each selected prompt $x_i$, a high quality answer $a_i$ is produced, for example by a larger model and/or human review.
    \item \textbf{Structure.} We store tuples $(x_i, a_i, v_i)$ where $v_i = \Phi(x_i)$. The set of all such entries is $C_{\text{static}}$.
\end{itemize}
Static hits are especially valuable: they serve curated answers with minimal latency and no new LLM inference.

\subsubsection{Dynamic tier $C_{\text{dynamic}}$}

The dynamic tier is a read write cache populated on the serving path.
\begin{itemize}
    \item \textbf{Population.} When a prompt cannot be served from the cache, the backend generates a response that is written into $C_{\text{dynamic}}$ together with its embedding.
    \item \textbf{Quality.} Entries reflect the response returned by the online agentic pipeline at that time. They are typically less vetted than static entries.
    \item \textbf{Eviction.} Standard cache policies such as LRU or TTL control the size of $C_{\text{dynamic}}$.
\end{itemize}

\paragraph{Why tiering improves quality.}
A tiered design separates responses by provenance and governance: the static tier can be populated only with curated, offline vetted answers and updated through controlled pipelines, while the dynamic tier can absorb fresh or long tail traffic under standard eviction and invalidation policies. Serving from the static tier when possible therefore improves consistency and reliability, and the dynamic tier provides coverage without relaxing the static tier's quality bar. Although we describe a two tier cache for concreteness, Krites only requires a curated (read only) tier backed by a mutable tier.

\subsubsection{Agentic backend $B$}

The backend $B$ is the agentic system (retrieval, tools, LLM decoding) that produces a fresh response when both tiers miss. Each call to $B$ pays the full inference cost that caching aims to avoid.

\subsection{Baseline GPTCache style policy}
\label{sec:baseline-policy}

Our baseline follows the static threshold policy used in GPTCache and similar systems \citep{bang2023gptcache,Li2024SCALM,schroeder2025vcache}. For a prompt $q$, the cache performs a nearest neighbor lookup in the static and dynamic tiers, then applies fixed similarity thresholds $\tau_{\text{static}}$ and $\tau_{\text{dynamic}}$ to decide whether to reuse or fall back to the backend.

The constants $\tau_{\text{static}}$ and $\tau_{\text{dynamic}}$ are tuned once and then applied uniformly across prompts. This is exactly the GPTCache style semantic cache: a nearest neighbor lookup followed by a fixed similarity cutoff that decides \emph{hit} versus \emph{miss}. Krites leaves this baseline policy unchanged on the serving path and only augments what happens after a static miss in the grey zone.

\begin{algorithm}[t]
\caption{Baseline GPTCache-style tiered semantic cache}
\label{alg:gptcache}
\begin{algorithmic}[1]
\Require prompt $q$
\State $v_q \gets \Phi(q)$ \Comment{Embed prompt}
\State $(h_{\text{static}}, s_{\text{static}}) \gets \textsc{NearestNeighbor}(C_{\text{static}}, v_q)$
\If{$s_{\text{static}} \ge \tau_{\text{static}}$}
    \State \Return $A(h_{\text{static}})$ \Comment{Static hit}
\EndIf
\State $(h_{\text{dynamic}}, s_{\text{dynamic}}) \gets \textsc{NearestNeighbor}(C_{\text{dynamic}}, v_q)$
\If{$s_{\text{dynamic}} \ge \tau_{\text{dynamic}}$}
    \State \Return $A(h_{\text{dynamic}})$ \Comment{Dynamic hit}
\EndIf
\State $a_{\text{gen}} \gets B(q)$ \Comment{Agentic backend}
\State Insert $(q, a_{\text{gen}}, v_q)$ into $C_{\text{dynamic}}$
\State \Return $a_{\text{gen}}$
\end{algorithmic}
\end{algorithm}

\section{Krites architecture}
\label{sec:Krites-architecture}

Krites augments the static-threshold baseline in Section~\ref{sec:baseline-policy} with an asynchronous verification loop. The serving path and thresholds remain unchanged: obvious hits and obvious misses are handled entirely by the static and dynamic tiers. The only difference is that when a prompt lies in a similarity grey zone with respect to the static tier, Krites schedules a background task that may later overwrite the dynamic entry with a higher quality static response via an \textbf{auxiliary overwrite}.

\begin{algorithm}[t]
\caption{Krites policy (non-blocking serving with asynchronous verification)}
\label{alg:krites-main}
\begin{algorithmic}[1]
\Require prompt $q$
\State $v_q \gets \Phi(q)$ \Comment{Embed prompt}
\State $(h_{\text{static}}, s_{\text{static}}) \gets \textsc{NearestNeighbor}(C_{\text{static}}, v_q)$
\If{$s_{\text{static}} \ge \tau_{\text{static}}$}
    \State \Return $A(h_{\text{static}})$ \Comment{Static hit, unchanged vs.\ baseline}
\EndIf
\State $(h_{\text{dynamic}}, s_{\text{dynamic}}) \gets \textsc{NearestNeighbor}(C_{\text{dynamic}}, v_q)$
\If{$s_{\text{dynamic}} \ge \tau_{\text{dynamic}}$}
    \State $a \gets A(h_{\text{dynamic}})$ \Comment{Dynamic hit}
\Else
    \State $a \gets B(q)$ \Comment{Agentic backend}
    \State Insert $(q, a, v_q)$ into $C_{\text{dynamic}}$ \Comment{Baseline write-back}
\EndIf
\If{$s_{\text{static}} \in [\sigma_{\min}, \tau_{\text{static}})$}
    \State enqueue asynchronous task \textsc{VerifyAndPromote}$(q, h_{\text{static}}, v_q)$
\EndIf
\State \Return $a$
\Statex
\Function{\textsc{VerifyAndPromote}}{$q, h_{\text{static}}, v_q$}
\State $a_{\text{static}} \gets A(h_{\text{static}})$
\State $\textit{approve} \gets J(q, h_{\text{static}}, a_{\text{static}})$
\If{$\textit{approve}$}
    \State upsert $(q, a_{\text{static}}, v_q)$ into $C_{\text{dynamic}}$ \Comment{Auxiliary overwrite}
\EndIf
\EndFunction
\end{algorithmic}
\end{algorithm}

The only additional work relative to Algorithm~\ref{alg:gptcache} is the grey zone test and the background call to \textsc{VerifyAndPromote}. All user facing decisions are still made by the original static thresholds, so Krites preserves the latency and on-path behavior of the baseline for the request that triggered verification.

\subsection{Grey-zone trigger and task scheduling}

We introduce a lower similarity cutoff $\sigma_{\min} < \tau_{\text{static}}$ that defines the grey zone for static candidates. Static neighbors with $s_{\text{static}} < \sigma_{\min}$ are treated as clearly non-interchangeable and are never sent to the judge, while candidates with $s_{\text{static}} \in [\sigma_{\min}, \tau_{\text{static}})$ are considered ambiguous and may be promoted if the judge confirms equivalence.

Operationally, \textsc{VerifyAndPromote} is executed by a background worker pool. In typical deployments, the worker pipeline includes (i) queueing and rate limiting, (ii) deduplication to avoid repeated judging of the same $(q, h_{\text{static}})$ pair, and (iii) retry logic with backoff for transient failures. Because the task is off path, queue depth affects only how quickly the pointer layer is populated, not serving latency.

\subsection{LLM judge requirements}

\paragraph{Task.}
The function $J(q,h,a)$ is a binary compatibility judge: given an incoming prompt $q$, a cached prompt $h$, and the cached answer $a$, it decides whether serving $a$ for $q$ is acceptable. In deployment, $J$ can be implemented with a strict, rubric-guided LLM prompt that checks intent alignment, entity/constraint consistency (names, locations, time, quantities), and freshness/personalization requirements, and returns a single-token decision (APPROVE/REJECT) at temperature~0.

\paragraph{Oracle for judge-agnostic policy evaluation.}
Our main trace-driven results (Section~\ref{sec:experiments}) are intentionally judge-agnostic: we instantiate $J$ as an oracle derived from the benchmark equivalence classes. This isolates the caching policy from judge prompt/model choices and avoids conflating judge behavior with occasional inconsistencies we observed in the benchmark IDs when manually auditing near-duplicate prompts.

\paragraph{Sanity checking judge realism.}
\label{judge check}
To ground practicality, we additionally evaluated a concrete LLM judge (Claude Opus~4.5) on a 100-pair human-audited sample drawn from the static similarity grey zone (i.e., pairs with $s_{\text{static}} \in [\sigma_{\min}, \tau_{\text{static}})$). Using the same strict rubric as above and a single-token output format, the judge agreed with human accept/reject labels on 99/100 pairs. This supports that high-precision LLM-based verification is feasible in practice, while Krites keeps verification off the critical path.




\subsection{Auxiliary overwrite semantics}

The auxiliary overwrite in \textsc{VerifyAndPromote} (Algorithm~\ref{alg:krites-main}) is what couples the static and dynamic tiers. Consider the dog-honey example in the introduction. The static cache contains a canonical prompt $h$ (``Can my dog have honey?'') with curated answer $A(h)$. A user issues $q$ (``What's the word on my dog having honey?''). If $s_{\text{static}} \in [\sigma_{\min}, \tau_{\text{static}})$, Krites enqueues \textsc{VerifyAndPromote}. If the judge approves, Krites upserts $(q, A(h), v_q)$ into $C_{\text{dynamic}}$. The next time $q$ (or a nearby paraphrase that hits this dynamic key) appears, the system serves the curated static answer from the dynamic tier.

In implementation, the overwrite should be idempotent and safe under concurrency. A common approach is to store a small amount of metadata with each dynamic entry (e.g., a ``static-origin'' bit and timestamp) and use last-writer-wins or timestamp guarded upserts to avoid overwriting a newer entry if that is undesirable in a particular deployment.

Krites does not modify the eviction policy of $C_{\text{dynamic}}$: promoted static entries are subject to the same LRU/TTL rules as all other dynamic entries. Verified pointers that are rarely used will eventually be evicted, and if the corresponding query reappears after eviction it again follows the baseline path and may trigger a new verification. This keeps the dynamic tier bounded without pinning static-derived entries and ensures that Krites inherits the capacity and freshness properties of the underlying dynamic cache.

\subsection{Cost and capacity considerations}

Krites introduces additional compute via off-path judge calls. The judge invocation rate is bounded by the frequency of grey-zone static misses and is controlled by $\sigma_{\min}$: raising $\sigma_{\min}$ reduces judge volume but also reduces recovered static hits, while lowering $\sigma_{\min}$ expands coverage but increases judge cost. Section~\ref{sec:discussion} discusses ROI tradeoffs and practical throttling strategies.

\section{Experimental evaluation}
\label{sec:experiments}

We evaluate Krites with trace-driven simulations on the two open benchmarks introduced by vCache \citep{schroeder2025vcache}. In all experiments we reuse the embeddings and the benchmark equivalence classes. Krites requires a binary decision $J(q,h,a)$ to approve promotions. To keep the evaluation policy-centric, judge-agnostic, and consistent with vCache, we do not run an LLM judge inside the simulation loop. Instead, we instantiate $J$ directly from the benchmark equivalence relation: we approve iff the query $q$ and the candidate neighbor $h$ share the same labeled equivalence class. To show the efficacy of current LLMs as judge we follow \ref{judge check}

\subsection{Datasets and static-dynamic split}
\label{subsec:split}

We use the following two datasets, which represent conversational and search style workloads. Their construction and statistics are described in detail in \citet{schroeder2025vcache}.
\begin{itemize}
    \item \textbf{SemCacheLMArena} ($\sim$60k prompts). A benchmark derived from Chatbot Arena logs with open ended conversational prompts and high lexical diversity.
    \item \textbf{SemCacheSearchQueries} ($\sim$150k prompts). A benchmark derived from the ORCAS query log with short, keyword-heavy search queries.
\end{itemize}

\paragraph{Why an oracle is a realistic proxy.}
The vCache benchmarks define interchangeability using LLM-as-a-judge labeling procedures, and SemCacheSearchQueries equivalence classes are constructed via a union-find pass guided by an LLM judge (GPT-4.1-nano). Using the benchmark equivalence relation as $J$ therefore evaluates Krites under the benchmark's own definition of correctness and provides an upper bound for any production verifier. However, in a small manual audit we observed occasional inconsistencies in the benchmark IDs for near-duplicate prompts, which motivates keeping our headline policy results judge agnostic rather than tying them to any single verifier implementation. To sanity check verifier realism, we additionally evaluated Claude Opus 4.5 as a strict binary judge on a 100 pair human-audited sample drawn from the static similarity grey zone, and found 99/100 agreement with human accept/reject labels.

\paragraph{History/evaluation split.}
To mimic deployment, we separate the data used to \emph{construct} the static tier from the data used to \emph{evaluate} online behavior. We first produce a deterministic request order by shuffling each benchmark once with a fixed seed. We then take the first 20\% of this ordered trace as a \emph{history} prefix and reserve the remaining 80\% as the \emph{evaluation} request stream processed in order during simulation. We report all hit/error metrics on the held out evaluation stream only (i.e., the history prefix is not used for evaluation), avoiding leakage from static construction into reported results.

\paragraph{Static tier construction (coverage based head selection).}
Using only the history prefix, we estimate the empirical popularity of each benchmark equivalence class. We then select the smallest set of equivalence classes whose cumulative frequency accounts for 60\% of history requests. The static tier contains one \emph{canonical} representative prompt per selected equivalence class, chosen deterministically as the shortest prompt in that class within the history prefix. Static entries are interpreted as precomputed answers available at time $t_0$.

\paragraph{Dynamic stream and fairness across policies.}
The remaining 80\% of requests form the dynamic traffic stream. The dynamic tier starts empty and is populated online as requests arrive (and, for Krites, may additionally receive verified promotions approved by $J$). Because both the baseline and Krites are evaluated on the same static preload, the same similarity threshold(s), and the same evaluation request order, this split affects the absolute headroom for static reuse but does not confound the relative comparison between policies.

\subsection{Baselines and configuration}

\paragraph{Static threshold baseline.}
Our baseline is the GPTCache-style policy in Algorithm~\ref{alg:gptcache}. For each dataset we set a single similarity threshold $t^\star$ and use $\tau_{\text{static}} = \tau_{\text{dynamic}} = t^\star$. Rather than tuning $t^\star$ ourselves, we reuse the best static thresholds identified in the vCache Pareto analysis: for each benchmark we choose the GPTCache configuration that lies on or near the static-threshold Pareto frontier at an error rate of roughly one to two percent (Figures 5 and 8 in \citet{schroeder2025vcache}). This ensures that our baseline represents a strong static policy under the same ground truth labeling.

\paragraph{Krites.}
Krites uses exactly the same embeddings, ground truth labels, and thresholds as the static baseline. The only difference is that it enables the asynchronous \textsc{VerifyAndPromote} loop from Algorithm~\ref{alg:krites-main} for static similarities in the grey zone $[\sigma_{\min}, \tau_{\text{static}})$. For our evaluations we take the $\sigma_{\min}$ as 0. All queries still traverse the baseline serving path (Algorithm~\ref{alg:gptcache}); the judge and auxiliary overwrite run off-path.

\subsection{Results}

Table~\ref{tab:static-hits} reports the fraction of requests served with a static-origin (curated) answer for the static baseline and Krites. For the baseline, this quantity is identical to the direct static tier hit rate; for Krites, it includes both direct static tier hits and dynamic tier hits created by auxiliary overwrite promotions.

\begin{table}[H]
\centering
\begin{tabular}{lccc}
\toprule
Dataset & Baseline & Krites & Relative gain \\
\midrule
SemCacheLMArena      & 8.2\%  & \textbf{19.4\%} & +136.5\% \\
SemCacheSearchQueries & 2.2\%  & \textbf{8.6\%} &  +290.3\% \\
\bottomrule
\end{tabular}
\caption{Static-origin served fraction for the tuned static baseline and Krites. For the baseline this equals the direct static tier hit rate; for Krites it includes both direct static tier hits and dynamic tier hits created by auxiliary overwrite promotions. The baseline thresholds are taken from the Pareto optimal GPTCache configurations reported in \citet{schroeder2025vcache}.}
\label{tab:static-hits}
\end{table}

\begin{figure}[t]
\centering
\begin{subfigure}[b]{0.40\linewidth}
  \centering
  \IfFileExists{template/a.png}{
    \includegraphics[width=\linewidth]{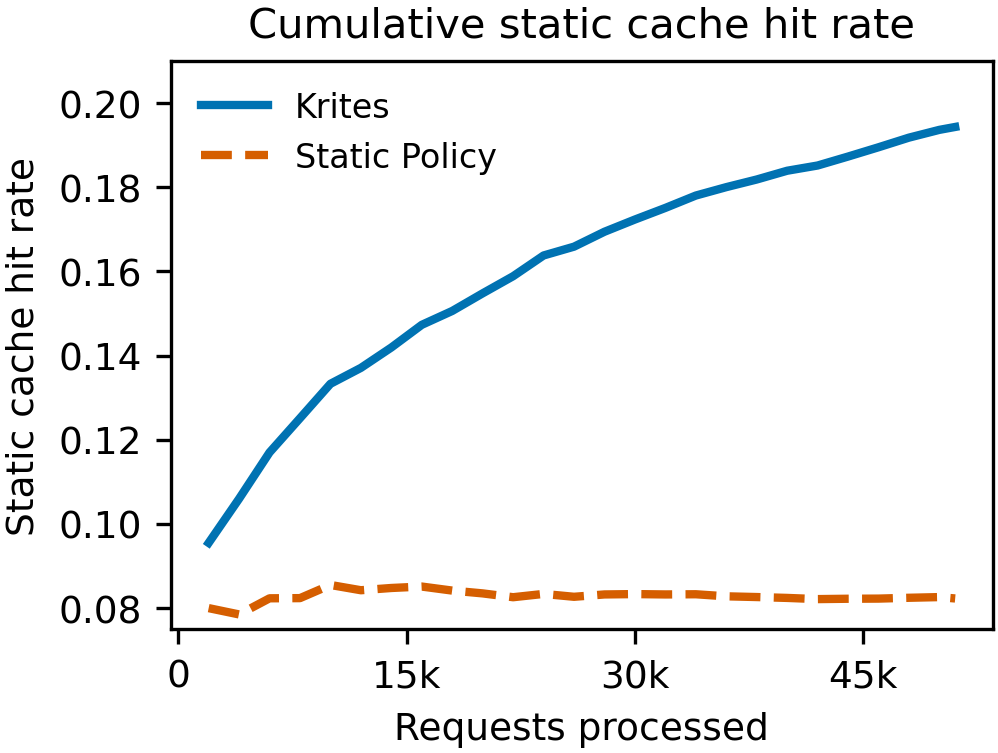}
  }{
    \fbox{\parbox{0.95\linewidth}{
      \centering
      \vspace{0.75em}
      \textbf{Missing file:} \texttt{template/a.png}
      \vspace{0.75em}
    }}
  }
  \caption{SemCacheLMArena static-origin served fraction over time.}
  \label{fig:static-hit-over-time-arena}
\end{subfigure}
\hfill
\begin{subfigure}[b]{0.40\linewidth}
  \centering
  \IfFileExists{template/b.png}{
    \includegraphics[width=\linewidth]{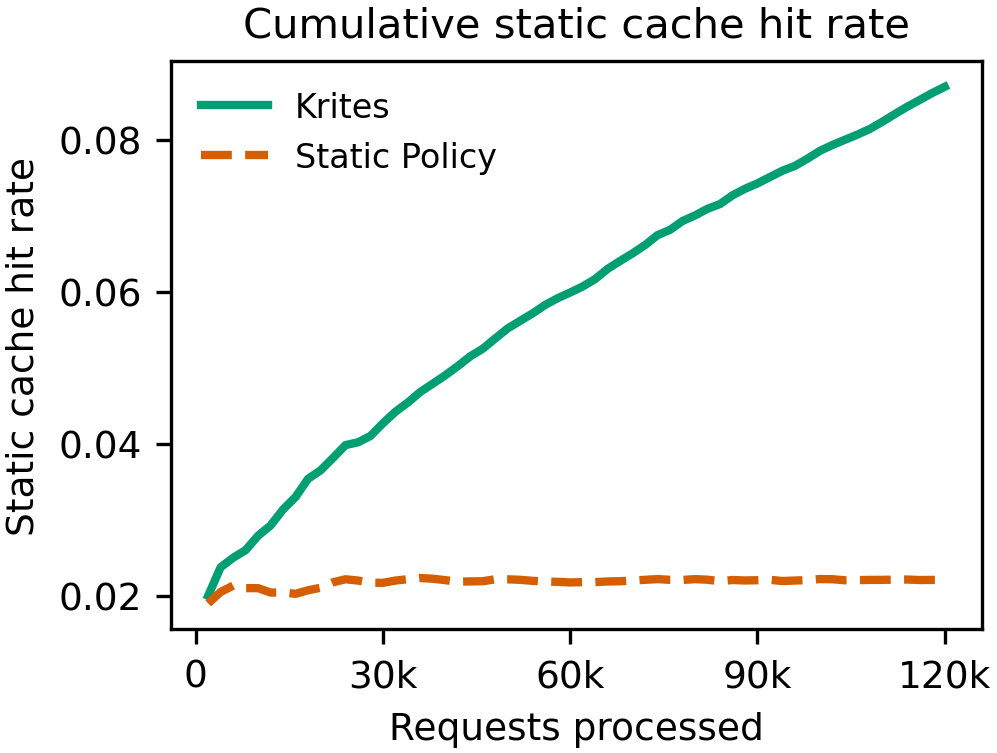}
  }{
    \fbox{\parbox{0.95\linewidth}{
      \centering
      \vspace{0.75em}
      \textbf{Missing file:} \texttt{template/b.png}
      \vspace{0.75em}
    }}
  }
  \caption{SemCacheSearchQueries static-origin served fraction over time.}
  \label{fig:static-hit-over-time-search}
\end{subfigure}

\caption{
Static-origin served fraction (direct static hits plus verified promotions) as a function of requests processed, starting from a cold dynamic cache, for (a) SemCacheLMArena and (b) SemCacheSearchQueries. Krites increases static-origin coverage over time by populating the dynamic tier with verified pointers to static answers via auxiliary overwrites.}
\Description{Two line plots showing the cumulative static-origin served fraction over processed requests for SemCacheLMArena and SemCacheSearchQueries, comparing Krites to a static threshold baseline.}
\label{fig:static-hit-over-time}
\end{figure}

On SemCacheLMArena, Krites more than doubles the fraction of traffic served with curated static answers. On SemCacheSearchQueries the gains are even bigger, where Krites increases static-origin serves by more than 290\% relative to the tuned static-threshold baseline, despite starting from thresholds already chosen to be Pareto-optimal in the vCache analysis \citep{schroeder2025vcache}.

Because Krites leaves the on path decision rule unchanged, the critical path latency and the error behavior of the baseline are preserved for the request that triggers verification.

\section{Discussion}
\label{sec:discussion}

Krites sits in a complementary space to prior work on semantic caching. Systems such as GPTCache and its variants apply static thresholds over embedding similarity in a single cache and have been widely adopted in practice \citep{bang2023gptcache,zilliz2023gptcachegithub,Li2024SCALM,Gill2024MeanCache}. vCache shows that this static design is fundamentally limited: similarity distributions for correct and incorrect candidates overlap heavily, and the optimal threshold varies widely across embeddings, so a single cutoff either violates error budgets or collapses toward exact-match behavior \citep{schroeder2025vcache}. Krites addresses a different question: given that operators often already run a tuned static-threshold policy in front of a tiered static-dynamic cache, and that static entries are curated and hard to update, how can we recover additional safe static hits \emph{without} changing the serving decision rule or its latency profile?

Several approaches attempt to improve the hit-error Pareto curve of semantic caching directly. One line fine tunes embedding models so that semantically equivalent prompts are closer and negatives are better separated \citep{zhu2024efficientembeddings}. Another learns embedding-specific thresholds and decision rules using online feedback and can provide user-defined error rate guarantees \citep{schroeder2025vcache}. These techniques significantly improve the hit rate versus error rate frontier for a single semantic cache. However, in many production deployments the static cache is effectively immutable or updated only by offline pipelines because its contents are tied to safety and quality review, which can leave curated responses stranded behind conservative global cutoffs.

\paragraph{Blocking verified caching.}
A natural alternative to Krites is a \textbf{blocking verified cache} that places an LLM judge directly in the serving path. In this design, borderline candidates (or even all cache hits) are synchronously checked by a judge before being served. Such a policy can increase cache hits by directly replacing a static threshold with a semantic equivalence test, but it adds an extra model call to many requests, increasing mean and tail latency and eroding the core benefit of caching for interactive workloads. Krites avoids this latency cost by decoupling serving from verification.

\paragraph{Assumption: verifier fidelity.}
Our trace-driven evaluation instantiates $J$ as an oracle defined by the vCache benchmark equivalence labels, so the headline gains characterize the policy under an ideal verifier for the benchmark notion of interchangeability. To assess realism, we additionally evaluated Claude Opus~4.5 as a strict binary judge on a 100-pair human-audited sample from the similarity grey zone and found 99/100 agreement with human accept/reject labels. In production, an LLM-based verifier may still have nonzero false reject and false approve rates. False rejects reduce promotion volume (reducing gains), while false approves can introduce errors on \emph{promoted} entries. Since verification is off path, such errors do not affect the request that triggered verification, but can affect subsequent hits to promoted keys; if promoted hit traffic is $p_{\text{prom}}$ and the verifier's false-approve rate on judged pairs is $\epsilon$, then the incremental error contribution from promotions is upper-bounded by $\epsilon\,p_{\text{prom}}$.

\subsection{Cost and ROI of off-path judging}

Krites introduces a new resource tradeoff: background judge compute in exchange for higher static-origin coverage. Let $\lambda$ be the request rate and let $p_{\text{grey}}$ be the fraction of requests whose nearest static neighbor falls in the grey zone $[\sigma_{\min}, \tau_{\text{static}})$. The raw judge request rate is approximately $\lambda_J \approx \lambda \, p_{\text{grey}}$ (before deduplication/rate limiting). Let $c_J$ denote the average compute cost per judge call.

The benefit of a single judge approval depends on (i) whether the pair is approved and (ii) how many times the promoted key is reused before eviction. Let $p_{\text{app}}$ be the approval probability and let $N$ be the number of subsequent requests that hit the promoted dynamic entry before eviction. Then the expected number of additional \emph{static-origin} serves attributable to one judge call is $\mathbb{E}[p_{\text{app}} N]$. This quantity is workload-dependent: it is larger when paraphrases recur and when the dynamic cache retains promoted entries long enough for reuse.

\paragraph{Model choice.}
Using a large model (e.g., Claude Opus~4.6) as the judge can be compute expensive because the judge prompt includes the cached answer, but it does not affect critical path latency. If compute budget is tight, operators can use a smaller judge model and/or gate judging on recurrence (e.g., only judge after observing a repeated query), trading off promotion volume and verifier precision/recall. In practice, the best verifier choice is application and workload dependent; the oracle-based evaluation in Section~\ref{sec:experiments} isolates Krites from these model-specific details.


%

\bibliographystyle{ACM-Reference-Format}
\bibliography{sample-base}

\end{document}